\documentclass[journal=jacsat,manuscript=special issue,14pt]{achemso}

\usepackage{amsfonts}
\usepackage{amsmath}
\usepackage{amssymb}
\usepackage{amsthm}
\usepackage{siunitx}
\usepackage{chemformula} 
\usepackage{datetime}
\usepackage{dcolumn} 
\usepackage{enumerate}
\usepackage[T1]{fontenc}
\usepackage{epigraph}
\usepackage{graphicx} 
\usepackage[utf8]{inputenc}
\usepackage[unicode]{hyperref}
\hypersetup{
    colorlinks=true,
    linkcolor=blue,
    filecolor=magenta,
    urlcolor=cyan,
}
\usepackage{mathptmx}
\usepackage{orcidlink}

\usepackage{tensor}
\usepackage{xcolor}
\usepackage{yfonts}

\newcommand{\R}{\mathbb{R}}  

\newcommand{\bfr}{{\bf r}}

\newcommand{\expect}[2]{\ensuremath{\langle #2 \vert #1 \vert #2 \rangle}}
\newcommand{\op}[1]{\mathsf #1}
\DeclareMathOperator{\dd}{d}
\DeclareMathOperator{\erf}{erf}
\DeclareMathOperator{\erfc}{erfc}
\newcommand{\mc}[1]{\multicolumn{1}{c}{#1}}

\makeatletter
\newcommand{\thickbar}{\mathpalette\@thickbar}
\newcommand{\@thickbar}[2]{{#1\mkern1.5mu\vbox{
  \sbox\z@{$#1\mkern-1.5mu#2\mkern-1.5mu$}%
  \sbox\tw@{$#1\overline{#2}$}%
  \dimen@=\dimexpr\ht\tw@-\ht\z@-.8\p@\relax
  \hrule\@height.8\p@ 
  \vskip\dimen@
  \box\z@}\mkern1.5mu}
}
\makeatother
\renewcommand*{\bar}{\thickbar}
\newcommand{\andreas}[1]{#1} 

\definecolor{maroon}{rgb}{0.5, 0., 0.}

\usepackage{achemso} 
\SectionNumbersOn    

\title{A modified expression for the Hamiltonian expectation value exploiting the short-range behavior of the wave function}

\author{Anthony Scemama\orcidlink{0000-0003-4955-7136}}
\affiliation{Laboratoire de Chimie et Physique Quantiques (UMR 5626),
Université de Toulouse, CNRS, UPS, France}
\email{scemama@irsamc.ups-tlse.fr}

\author{Andreas Savin\orcidlink{0000-0001-8401-8037}}
\affiliation{Laboratoire de Chimie Th\'eorique, CNRS and Sorbonne University \\ 4 place Jussieu, 75252 Paris, France}
\email{andreas.savin@lct.jussieu.fr}


\begin{document}


\begin{abstract}
The expectation value of the Hamiltonian using a model wave function is widely used to estimate the eigenvalues of electronic Hamiltonians.
We explore here a modified formula for models based on long-range interaction.
It scales differently the singlet and triplet component of the repulsion between electrons not present in the model (its short-range part).
The scaling factors depend uniquely on the parameter used in defining the model interaction, and are constructed using only exact properties.
We show results for the ground states and low-lying excited states of Harmonium with two to six electrons.
We obtain important improvements for the estimation of the exact energy, not only over the model energy, but also over the expectation value of the Hamiltonian.
\end{abstract}

\maketitle

\newpage
\tableofcontents
\newpage

\centerline{\today \hspace{2pt} at \currenttime}

\section{Introduction}
\label{introduction}

A widespread procedure to estimate the energy of a given state is to use an approximate normalized wave function $\Psi(\mu)$, and compute the expectation value of the Hamiltonian  of the system of interest, $\op{H}$,
\begin{equation}
    \label{eq:var}
    E_\op{H}(\mu)=\expect{\op{H}}{\Psi(\mu)}.
\end{equation}

When $\Psi(\mu)$ is constructed from orbitals, it is difficult to obtain a good description of the short-range behavior of the wave function.~\cite{Gil-63}
This is associated to the singularity of the Coulomb repulsion between electrons when their separation vanishes.~\cite{KutMor-JCP-92}
This paper uses a model wave function, $\Psi(\mu)$, produced by a long-range model.
To improve over $E_\op{H}(\mu)$, Eq.~\eqref{eq:var}, we use an adiabatic connection, and information about the exact short-range behavior of the wave function.

We first recapitulate the basic principles of our method~\cite{Sav-JCP-20, SavKar-JCP-23}.
In previous papers, we presented results for only $N=2$ electrons.
Here, we extend the method to a larger number of electrons.
We obtain the ground state energies and those of the lowest excited state for electrons confined by a quadratic potential (Harmonium), which we can use for comparison with accurate energies
published in the literature.

\section{Method}
\label{method}
\subsection{Model system and an adiabatic connection}
We consider $\Psi(\mu)$ to be the eigenfunction of a model system with Hamiltonian $\op{H}(\mu)$,
\begin{equation}
    \label{eq:model-se}
    \op{H}(\mu) \Psi(\mu) = E(\mu) \Psi(\mu),
\end{equation}
where $\Psi(\mu)$ is normalized, and the Coulomb repulsion between electrons showing up in $\op{H}(\mu)$ is replaced by
\begin{equation}
    \label{eq:W}
    \op{W}(\mu) = \sum_{1 \le i < j \le N} w(|\bfr_i - \bfr_j|, \mu)
\end{equation}
where $w(0,\mu)$ is finite for every $\mu < \infty$.
Within the scope of this paper,
\begin{equation}
    \label{eq:w}
    w(r,\mu) = \frac{\erf(\mu r)}{r} .
\end{equation}
It is noteworthy that $w(r,0)=0$ and asymptotically, $w(r,\infty)=1/r$.

Even if solved accurately, $E(\mu)$ has no physical significance.
To obtain the physical energy, $E$, one must estimate the difference $E-E(\mu)$.
For example, if we use Eq.~\eqref{eq:var},  this difference can be approximated
by the expectation value of $\bar{\op{W}}(\mu)$. One approximates
\begin{equation}
    \label{eq:var-wbar}
    E \approx E_\op{H}(\mu) = E(\mu) + \expect{\bar{\op{W}}(\mu)}{\Psi(\mu)}
\end{equation}
where, using Eq.~\eqref{eq:W},
\begin{equation}
    \label{eq:Wbar}
    \bar{\op{W}}(\mu) = \sum_{1 \le i < j \le N} \bar{w}(|\bfr_i - \bfr_j|, \mu) ,
\end{equation}
and
\begin{equation}
    \label{eq:wbar}
    \bar{w}(r,\mu) = \frac{\erfc(\mu r)}{r} .
\end{equation}

In this paper, we try to go beyond Eq.~\eqref{eq:var} without an  exaggerated computational effort.~\cite{Sav-JCP-20}

In principle, the exact energy, $E$, an eigenvalue of $\op{H}=\op{H}(\mu=\infty)$, can be obtained through an adiabatic connection.
For it, we define the Hamiltonian
\begin{equation}
    \label{eq:ac-H}
    \op{H}(\lambda,\mu) = \op{H}(\mu) + \lambda  \bar{\op{W}}(\mu) .
\end{equation}
The corresponding Schr\"odinger equation is
\begin{equation}
    \label{eq:ac-se}
    \op{H}(\lambda,\mu) \Psi(\lambda,\mu) = E(\lambda, \mu) \, \Psi(\lambda,\mu).
\end{equation}
Note that $\op{H}=\op{H}(\lambda=1,\mu)=\op{H}(\lambda,\mu=\infty)$ and $\op{H}(\mu)=\op{H}(\lambda=0,\mu)$.
Similar omissions of $\lambda$ or/and $\mu$ in the notations are applied to $E$ and $\Psi$.
By using the Hellmann-Feynman theorem, one has
\andreas{
\begin{equation}
    \label{eq:acE}
    E = E(\mu) + \int_0^1 d\lambda \frac{\partial E(\lambda,\mu)}{\partial \lambda} =
    E(\mu) + \int_0^1 \dd \lambda \, \expect{\bar{\op{W}}(\mu)}{\Psi(\lambda,\mu)} .
\end{equation}
}

In contrast to $E_\op{H}$, Eq.~\eqref{eq:var}, Eq.~\eqref{eq:acE} is in principle exact, but requires the knowledge of $\Psi(\lambda,\mu)$ for all $\lambda$.
However, we assume that we know only the result for the wave function of the model system, $\Psi(\mu)=\Psi(\lambda=0,\mu)$.
In order to deal with this problem we exploit the short-range nature of the operator $\bar{\op{W}}$, Eq.~\eqref{eq:Wbar}: for obtaining $\expect{\bar{\op{W}}(\mu)}{\Psi(\lambda,\mu)}$ only the short-range part of the wave function is needed, and we can benefit from some exact, universal knowledge about the latter (see, e.g., ref.~\citenum{KurNakNak-AQC-16}).

While this approach avoids the Coulomb singularity at electron-electron coalescence for any finite $\mu$, it is nevertheless computationally preferable to have a weak interaction, that is, a small value of $\mu$.
Unfortunately, the solution we propose becomes exact only for large $\mu$.
As this model worsens considerably when $\mu$ is small, the common practice would be to use a mean field potential when the interaction between electrons is turned off.
However, findings from a previous study~\cite{SceSav-JCC-24} indicate that resorting to a mean field potential has no noticeable impact on the corrections applied, as long as the results are within the bounds of chemical accuracy.
This can be understood by considering the model's focus on the short-range behavior of the wave function.
In situations where electrons are close, the repulsive force between them is the dominant interaction, overshadowing the effects of the one-particle potential.
Furthermore, it is worth noting that the minimal error introduced by the mean field approximation at $\mu=0$, characterized at the Hartree-Fock level, is essentially the correlation energy.
The magnitude of this error is usually significantly larger, by at least an order of magnitude, than the threshold defined by
{\em chemical accuracy}~\cite{Pop-RMP-99}, $\pm 1$~kcal/mol~$\approx$\SI{1.6}{\milli \hartree}.

We explore how small we can choose $\mu$ without sacrificing accuracy, declaring ourselves satisfied if the errors are smaller than the chemical accuracy.

\subsection{Asymptotic short-range behavior}
The behavior $\Psi(\lambda,\mu)$ for small $r=|\bfr_1-\bfr_2|$ and large $\mu$ is known~\cite{GorSav-PRA-06,SavKar-JCP-23},
\begin{equation}
    \label{eq:psi-asy}
    \Psi(\lambda,\mu) = \sum_{\ell,m} \mathcal{N}_{\ell,m} \, \varphi_\ell(r;\lambda,\mu)\, Y_{\ell,m}(\Omega),  \qquad \text{for } r \rightarrow 0, \mu \rightarrow \infty.
\end{equation}
$\mathcal{N}_{\ell,m}$ depends on all space and spin variables, except $\bfr=\bfr_1-\bfr_2$.
In particular, it does not depend on $\lambda$ and $\mu$; it is related to the exact wave function ($\mu=\infty)$.~\cite{GorSav-PRA-06}
$Y_{\ell,m}$ are the spherical harmonics, $\Omega$ is the solid angle associated with $\bfr=\bfr_1-\bfr_2$,
\andreas{
not to be confused with that associated to $\bfr_i$ in systems with spherical symmetry. Furthermore,
}
\begin{align}
\label{eq:varphi-asy}
\varphi_\ell(r;\lambda,\mu)\,\propto\,&r^\ell\left[1+\frac{\lambda\,r}{2\ell+2}
+\frac{1-\lambda}{2\ell+2}\left(r\,\erf(\mu r)+\frac{2\ell+2}{2\ell+1}
\frac{e^{-\mu^2 r^2}}{\mu\sqrt{\pi}}\right.\right.\\
& \nonumber
+\left.\left.\frac{\Gamma(\ell+3/2)-\Gamma(\ell+3/2,\mu^2r^2)}
{\sqrt{\pi}(2\ell+1)\mu^{2\ell+2}r^{2\ell+1}}\right)\right]
\end{align}
The incomplete gamma function is:
\[ \Gamma(a,z) = \int_z^\infty \dd t \; t^{a-1} e^{-t} \]
and $\Gamma(a)=\Gamma(a,0)$.
Eq.~\eqref{eq:psi-asy} satisfies Kato's cusp condition~\cite{Kat-CPAM-57} for $\mu=\infty$.

For the integrand appearing in the adiabatic connection, Eq.~\eqref{eq:acE}, it is useful to define the second order reduced density matrix,
\begin{equation}
    \label{eq:def-P2}
    P_2(\bfr_1,\bfr_2;\bfr_1',\bfr_2';\lambda,\mu) = N(N-1) \sum_{\sigma_1, \dots, \sigma_N} \int_{\R^{3N-6}} \dd  \bfr_3 \dots \dd \bfr_N \,\Psi(\bfr_1, \dots, \bfr_N; \mu) \, \Psi^*(\bfr_1', \dots, \bfr_N';\mu).
\end{equation}
as it gives
\begin{equation}
    \label{eq:wbar-p2}
    \expect{\bar{\op{W}}(\mu)}{\Psi(\lambda,\mu)} = \frac{1}{2} \int_{\R^6} \dd \bfr_1 \dd \bfr_2 \, P_2(\bfr_1,\bfr_2;\bfr_1,\bfr_2;\lambda,\mu)\,\bar{w}(|\bfr_1-\bfr_2|; \lambda, \mu) .
\end{equation}
For large $\mu$, we can use Eq.~\eqref{eq:psi-asy}, and Eq~\eqref{eq:def-P2}, to get the asymptotic ($\mu \rightarrow \infty$) expression
\begin{equation}
    P_{2,\text{asy}}(\bfr_1,\bfr_2;\bfr_1',\bfr_2';\lambda,\mu) =\label{eq:p2-asy}
    \sum_{\ell,m} \sum_{\ell',m'} \mathcal{C}_{\ell,m,\ell',m'} \, \varphi_\ell(r;\lambda,\mu) \varphi_{\ell'}(r;\lambda,\mu) Y_{\ell,m}(\Omega) Y_{\ell',m'}^*(\Omega')
\end{equation}
$\mathcal{C}_{\ell,m,\ell',m'}$ is produced by the integration over all variables except $\bfr_1-\bfr_2$.
Note that
\begin{equation}
    \label{eq:permut-p2}
    P_{2,\text{asy}}(\bfr_1,\bfr_2;\bfr_1',\bfr_2';\lambda,\mu) = (-1)^\ell P_{2,\text{asy}}(\bfr_1,\bfr_2;\bfr_2',\bfr_1';\lambda,\mu).
\end{equation}
For obtaining the expectation value of $\bar{\op{W}}(\mu)$, Eq.~\eqref{eq:wbar-p2}, we also integrate over the angular variables, and use the orthogonality of the $Y_{\ell,m}$,
\begin{equation}
    \label{eq:wbar-p2-asy}
    \expect{\bar{\op{W}}(\mu)}{\Psi(\lambda,\mu)}  = \sum_{\ell=0} c_\ell \mathcal{I}_\ell(\lambda,\mu)
\end{equation}
where
\begin{equation}
    \label{eq:int-asy}
    \mathcal{I}_\ell (\lambda,\mu)  =  \int_{\R} \dd r\,  r^2 \, |\varphi_\ell(r;\lambda,\mu)|^2\, \bar{w}(r,\mu).
\end{equation}
The coefficients $c_\ell$ come from the integration of $\sum_m \mathcal{C}_{\ell,m,\ell',m'} Y_{\ell,m}(\Omega) Y_{\ell,m}(\Omega)$ over the angular variables, and are independent of $\lambda$ and $\mu$.
Note that integration over $\Omega$ and the orthogonality condition of the $Y_{\ell,m}$ eliminates the dependence on $\ell', m'$.
The one-dimensional integrals, Eq.~\eqref{eq:int-asy}, have been computed once and for all (see Eq. (25) and (26) of ref.~\citenum{SavKar-JCP-23}).

The adiabatic connection, Eq.~\eqref{eq:acE}, requires the integration of Eq.~\eqref{eq:wbar-p2-asy} over $\lambda$.
The integral of $\mathcal{I}_\ell(\lambda,\mu)$, on the r.h.s. of Eq.~\eqref{eq:wbar-p2-asy}, can be performed analytically.~\cite{Sav-JCP-20,SavKar-JCP-23}
One has still to determine the coefficients $c_\ell$.
To deal with this problem we choose, in this paper, to truncate the sum over $\ell$ to some maximal value $L$.
To justify it, we note that, for small $r$, $\varphi_\ell$ is of order $r^{\ell}$.
Thus, for a given short-range interaction, $\bar{\op{w}}(\mu)$, $\mathcal{I}_\ell(\lambda, \mu)$ vanishes faster with increasing $\ell$.
By a change of the integration variable from $r$ to $\mu r$ in $\mathcal{I}_\ell(\lambda, \mu)$, Eq.~\eqref{eq:int-asy}, we find that the leading term of the expansion in $1/\mu$ is of order $\mu^{-(2\ell +2)}$.


The strongest cutoff in Eq.~\eqref{eq:wbar-p2-asy} corresponds to choosing $L=0$,
\begin{equation}
    \label{eq:L=0}
    \expect{\bar{\op{W}}(\mu)}{\Psi(\mu)} \approx  c_0 \mathcal{I}_0(0,\mu).
\end{equation}
We solve Eq.~\eqref{eq:L=0} for $c_0$.
The only integral needed is that on the l.h.s. of Eq.~\eqref{eq:L=0}; it is already used in the computation of the expectation value of $\op{H}$, Eq.~\eqref{eq:var-wbar}.

For the less severe cutoff, $L=1$, we first separate $P_2$ into a {\em singlet} and a {\em triplet}  part~\cite{McWKut-ICQC-68, Ten-JCP-04},
\begin{align}
    P_s(\bfr_1,\bfr_2;\bfr_1',\bfr_2';\lambda,\mu) & = \frac{1}{2}\left( P_2(\bfr_1,\bfr_2;\bfr_1',\bfr_2';\lambda,\mu) +     P_2(\bfr_1,\bfr_2;\bfr_2',\bfr_1';\lambda,\mu) \right) \\
    P_t(\bfr_1,\bfr_2;\bfr_1',\bfr_2';\lambda,\mu) & = \frac{1}{2}\left( P_2(\bfr_1,\bfr_2;\bfr_1',\bfr_2';\lambda,\mu) - P_2(\bfr_1,\bfr_2;\bfr_2',\bfr_1';\lambda,\mu) \right) .
\end{align}
This defines
\begin{align}
    \label{eq:wbars-ident}
    \expect{\bar{\op{W}}(\mu)}{\Psi(\lambda,\mu)}_s & = \frac{1}{2} \int_{\R^6} \dd \bfr_1 \dd \bfr_2 \, P_s(\bfr_1,\bfr_2;\bfr_1,\bfr_2;\lambda,\mu)\,\bar{w}(|\bfr_1-\bfr_2|,\mu) , \\
    \label{eq:wbart-ident}
    \expect{\bar{\op{W}}(\mu)}{\Psi(\lambda,\mu)}_t & = \frac{1}{2} \int_{\R^6} \dd \bfr_1 \dd \bfr_2 \, P_t(\bfr_1,\bfr_2;\bfr_1,\bfr_2;\lambda,\mu)\,\bar{w}(|\bfr_1-\bfr_2|,\mu).
\end{align}
We can apply this also to the asymptotic expressions for the two-body density matrix, $P_{2,\text{asy}}$, Eq.~\eqref{eq:p2-asy}.
When using Eq.~\eqref{eq:permut-p2} for $L=1$, $P_{s,\text{asy}}$ retains the $\ell=0$ (singlet) component and $P_{t,\text{asy}}$ retains the $\ell=1$ (triplet) component,
\begin{align}
   \label{eq:L+1,0}
    \frac{1}{2} \int_{\R^6} \dd \bfr_1 \dd \bfr_2 \, P_s(\bfr_1,\bfr_2;\bfr_1,\bfr_2;\lambda=0,\mu)\bar{w}(|\bfr_1-\bfr_2|,\mu)
     & \approx c_0 \mathcal{I}_0(0,\mu)  \\
    \label{eq:L+1,1}
    \frac{1}{2} \int_{\R^6} \dd \bfr_1 \dd \bfr_2 \,P_t(\bfr_1,\bfr_2;\bfr_1,\bfr_2;\lambda=0,\mu)\bar{w}(|\bfr_1-\bfr_2|,\mu) & \approx  c_1  \mathcal{I}_1(0,\mu).
\end{align}
So, we obtain the approximation
\begin{align}
    \label{eq:wbars}
    \expect{\bar{\op{W}}(\mu)}{\Psi(\lambda=0,\mu)}_s &  \approx c_0 \mathcal{I}_0(0,\mu) , \\
    \label{eq:wbart}
    \expect{\bar{\op{W}}(\mu)}{\Psi(\lambda=0,\mu)}_t & \approx c_1 \mathcal{I}_1(0,\mu).
\end{align}

Summarizing, the formulas used in this paper are
\begin{align}
    \label{eq:ac-0}
    E & \approx E_{L=0}(\mu) = E(\mu) + \alpha_0(\mu) \expect{\bar{\op{W}}(\mu)}{\Psi(\mu)} \\
    \label{eq:ac-1}
    E & \approx E_{L=1}(\mu) = E(\mu) +\alpha_0(\mu) \expect{\bar{\op{W}}(\mu)}{\Psi(\mu)}_s + \alpha_1(\mu) \expect{\bar{\op{W}}(\mu)}{\Psi(\mu)}_t ,
\end{align}
where we have introduced the prefactors $\alpha_0(\mu)$ and $\alpha_1(\mu)$,
\begin{equation}
    \label{eq:alpha}
    \alpha_\ell(\mu) = \frac{ \int_0^1 \dd \lambda \, \mathcal{I}_\ell(\lambda,\mu)}{\mathcal{I}_\ell(0,\mu)}.
\end{equation}
Explicit expressions for $\alpha_\ell$ are~\cite{SavKar-JCP-23},
\begin{align}
    \label{eq:alpha-0-mu}
    \alpha_0 (\mu) & = \frac{0.319820 + 1.063846 \mu + \mu^2}{0.487806 + 1.375439 \mu + \mu^2} \\
    \label{eq:alpha-1-mu}
    \alpha_1 (\mu) & = \frac{0.113074 + 0.638308 \mu + \mu^2}{0.122652 + 0.674813 \mu + \mu^2}
\end{align}

Note that we can also add and subtract $\expect{\bar{\op{W}}(\mu)}{\Psi(\mu)}$ from the expressions of $E_{L=0}$, Eq.~\eqref{eq:ac-0}, or $E_{L=1}$, Eq.~\eqref{eq:ac-1}.
This allows to express the same approximations as corrections to $E_\op{H}(\mu)$,
\begin{align}
    \label{eq:ac-0-h}
    E_{L=0}(\mu) & = E_\op{H}(\mu) + \left[\alpha_0(\mu) -1 \right] \expect{\bar{\op{W}}(\mu)}{\Psi(\mu)} \\
    \label{eq:ac-1-h}
    E_{L=1}(\mu) & = E_\op{H}(\mu) +\left[\alpha_0(\mu) -1 \right] \expect{\bar{\op{W}}(\mu)}{\Psi(\mu)}_s + \left[\alpha_1(\mu) -1 \right] \expect{\bar{\op{W}}(\mu)}{\Psi(\mu)}_t ,
\end{align}
As
\begin{equation}
 \label{eq:alpha-ineq}
    \alpha_0(\mu) \le \alpha_1(\mu) \le 1
\end{equation},
\begin{equation}
   \label{eq:e-ineq}
   E_{L=0}(\mu) \le E_{L=1}(\mu) \le E_\op{H}(\mu).
\end{equation}

It is important to emphasize that the parameters $\alpha_\ell(\mu)$ are solely dependent on the value of $\mu$.
Nevertheless, the validity range of $\mu$ for the asymptotic approximation depends on the system or the electronic state under study.

Furthermore, we would like to draw the attention to an inconsistency in our approximation.
The expression of $\alpha_0 (\mu \rightarrow \infty)$ is correct up to order $\mu^{-3}$.
For $\alpha_1 (\mu \rightarrow \infty)$, the coefficients of the expansion in $1/\mu$ are zero for up to order $\mu^{-3}$,  and we take into account corrections of order $\mu^{-4}$ and $\mu^{-5}$, orders that are not exactly described by $\alpha_0$.
Going beyond the order $\mu^{-3}$ is possible for $\alpha_0$ is feasible, but is not treated in the present paper.

\section{Numerical results}
\label{numericalresults}
\subsection{Computational details}
We consider $N$ electrons confined by a harmonic potential,
\begin{equation}
    \label{eq:V}
    \op{V} = \frac{1}{2} \omega^2 \sum_{i=1,N} \bfr_i^2
\end{equation}
and the potential $W(\mu)$, Eq.~\eqref{eq:W}, for all electron-electron interactions.
We present here results obtained for $\omega=1/2$.

Recall that in  the non-interacting system, that is the harmonic oscillator, the orbital energy ordering is given by $n+2\ell$, where $n=0, 1, \dots$.
Within this framework, the minimum energy configuration is attributed to an $s$-state ($n=0$), highlighting the principle that the $p$ orbitals ($n=0, \ell=1$) are filled prior to the occupancy of the first excited $s$-state ($n=1, \ell=0$).
Consequently, for a system with $N=3$ electrons, the ground state configuration diverges from a $^2S$ symmetry, instead manifesting a $^3P$ symmetry.

Calculating the expectation values $\expect{\bar{\op{W}}(\mu)}{\Psi(\lambda=0,\mu)}_s$ and $\expect{\bar{\op{W}}(\mu)}{\Psi(\lambda=0,\mu)}_t$, as required in Eqs.~\eqref{eq:ac-0} and \eqref{eq:ac-1}, is rather straightforward within conventional codes.
In a basis of orbitals $\phi_i$, and with $P_2$ normalized to $N(N-1)$, the two-particle density matrix in this basis can be expressed as
\begin{equation}
P_2(\bfr_1,\bfr_2;\bfr_1',\bfr_2';\lambda=0,\mu) = \sum_{i,j,k,l} P_{i,j,k,l}(\lambda=0,\mu)\,\phi_i^*(\bfr_1') \phi_j^*(\bfr_2') \phi_k(\bfr_1) \phi_l(\bfr_2),
\end{equation}
leading to the following formulations for the expectation values for singlet and triplet contributions:
\begin{align}
 \expect{\bar{\op{W}}(\mu)}{\Psi(\mu)}_{s,t}
   & = \frac{1}{4} \sum_{i,j,k,l}\left[ P_{i,j,k,l}(0,\mu) \pm P_{i,j,l,k}(0,\mu) \right] \langle i j | k l \rangle_\mu
  \\
  & =  \frac{1}{4} \sum_{i,j,k,l}P_{i,j,k,l}(0,\mu) \left( \langle i j | k l \rangle_\mu \pm \langle i j | l k\rangle_\mu \right),
\end{align}
where the integrals
\begin{equation}
    \langle i j| k l \rangle_\mu = \int_{\R^6} \dd \bfr_1 \dd \bfr_2 \, \phi_i^*(\bfr_1) \phi_j^* (\bfr_2) w(|\bfr_1-\bfr_2|,\mu) \phi_k(\bfr_1) \phi_l (\bfr_2).
\end{equation}
represent the electron-electron repulsion integrals within the modified potential $w(r,\mu)$.

Eqs.~\eqref{eq:ac-0} and \eqref{eq:ac-1} were implemented in the \textsc{Quantum Package}\cite{garniron_2019} program,
which served as the computational platform for our calculations.

\begin{table}[]
    \centering
    \begin{tabular}{c S[table-format=1.10]}
    \hline
        \mc{Angular momentum}  & \mc{Exponent} \\
            \hline
      $s$  & 0.0395061728 \\
           & 0.0592592593 \\
           & 0.0888888889 \\
           & 0.1333333333 \\
           & 0.2          \\
           & 0.3          \\
           & 0.45         \\
           & 0.675        \\
           & 1.0125       \\
        \hline
      $p$  & 0.0888888889 \\
           & 0.1333333333 \\
           & 0.2          \\
           & 0.3          \\
           & 0.45         \\
           & 0.675        \\
           & 1.0125       \\
        \hline
      $d$  & 0.1777777778 \\
           & 0.2666666667 \\
           & 0.4          \\
           & 0.6          \\
           & 0.9          \\
        \hline
      $f$  & 0.3333333333 \\
           & 0.5          \\
           & 0.75         \\
        \hline
      $g$  & 0.4          \\
           & 0.6          \\
           & 0.9          \\
        \hline
      $h$  & 0.7          \\
        \hline
      $i$  & 0.8          \\
    \hline
    \end{tabular}
    \caption{Exponents of the 9s7p5d3f3g1h1i Gaussian basis set.}
    \label{tab:basis}
\end{table}

In this paper we aim only to study the effect of our approximation.
Thus, we have not made any attempt to reduce the computational cost of the calculation.
In the computational studies detailed herein, our objective has been to achieve the full configuration interaction (FCI) level of accuracy utilizing a 9s7p5d3f3g1h1i even-tempered Gaussian basis set as shown in Table~\ref{tab:basis}.
For $N=2$, FCI calculations were performed for all values $\mu$.
For systems with ($2 < N \le 6$) electrons, we employed the Configuration Interaction using a Perturbative Selection made Iteratively (CIPSI) algorithm,\cite{huron_1973,garniron_2019} aiming to attain the highest feasible approximation to FCI while maintaining a minimized computational cost.

The adopted computational protocol involved initially performing a CIPSI calculation with $\mu=\infty$, using Hartree-Fock orbitals.
The determinant selection continued until the second-order perturbative correction ($E_{\text{PT2}}$) decreased to less than \SI{1}{\milli\hartree}, or until the number of selected determinants exceeded 100~000.
The natural orbitals derived from this wave function were used in a subsequent CIPSI calculation, until $E_{\text{PT2}}$ was reduced to less than \SI{0.1}{\milli\hartree}, or the determinant count surpassed the 10 million threshold.
This final computation yields three critical outcomes: the extrapolated FCI energy ($E_{\text{exFCI}}$), a basis of Slater determinants in which a  wavefunction nearly equivalent to the FCI wave function is expanded, and its associated variational energy $E_{\text{CI}}$.

The choice of different $\mu$ values leads to distinct sets of Hartree-Fock or natural orbitals.
To obtain the best possible accuracy, a new calculation should be made from the start when the value of $\mu$ is changed.
To reduce the computational cost of our calculations, we take advantage of the fact that the FCI wave function can be expressed using any set of orbitals.
As we are close to the FCI, we can keep the orbitals and the set of selected determinants fixed for all values of $\mu$, and obtain the wave function $\Psi(\mu)$ by diagonalizing $H(\mu)$ in the basis of selected determinants obtained at $\mu = \infty$. We choose the particular value of $\mu=\infty$ as it is the value that generates the largest set of determinants with non-negligible coefficients.


\subsection{Checking the approximation of the scaling factors \texorpdfstring{$\alpha(\mu)$}{α(μ)}}
\begin{figure}
    \centering
    \includegraphics[width=0.8\columnwidth]{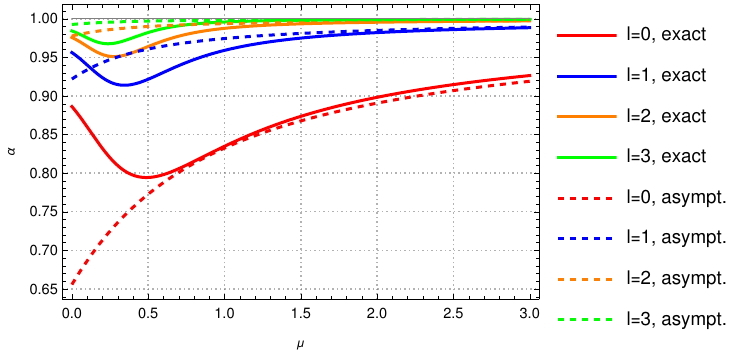}
    \caption{The expressions of the prefactors of  $\alpha_\ell$, as functions of $\mu$; exact, Eq.~\eqref{eq:alpha-exact}, full curves, and asymptotically correct, Eq.~\eqref{eq:alpha}, dashed curves; for $\ell=0$ (red), $\ell=1$ (blue), $\ell=2$ (orange), $\ell=3$ (green). The horizontal gray line shows the value that is chosen for obtaining the expectation value of  $\op{H}$.}
    \label{fig:alphas}
\end{figure}
We start by analyzing some results for a system with $N=2$ electrons.
In this system, the equation for $\bfr_1-\bfr_2$ can be separated.
Furthermore, as the potential depends only on $r$, the wave functions are eigenfunctions of the angular momentum, and only one term (with the corresponding $\ell$) remains on the r.h.s. of Eq.~\eqref{eq:wbar-p2-asy}.
Thus, the prefactors of $\expect{\bar{\op{W}}(\mu)}{\Psi(\mu)}$ that reproduce the exact energies are given by
\begin{equation}
    \label{eq:alpha-exact}
    \alpha_{\ell,\text{exact}} (\mu) = \frac{E- E(\mu)}{\expect{\bar{\op{W}}(\mu)}{\Psi(\mu)}}
\end{equation}
(The dependence of $\alpha$ on $\ell$ appears here through the dependence of $\Psi$ on $\ell$.)
Fig.~\ref{fig:alphas} compares $\alpha_{\ell,\text{exact}}$ with the approximations $\alpha_\ell$ of Eq.~\eqref{eq:alpha} for the lowest energy states for $\ell=0,1,2,3$.
$\ell=0, \text{ or }2$ correspond to singlets, $\ell=1, \text{ or }3$ correspond to triplets ($\ell=2, \text{ or }3$ are {\em non-natural} according to the terminology of ref.~\citenum{KutMor-JCP-92}).
Of course, $\alpha_\ell(\mu) \rightarrow \alpha_{\ell,\text{exact}}(\mu)$ as $\mu$ increases.
As $\ell$ increases $\alpha_{\ell,\text{exact}}(\mu), \alpha_\ell(\mu)$ approach the constant value 1 corresponding to $\expect{\op{H}}{\Psi(\mu)}$.
However, as $\mu$ decreases, there is a change of the behavior of $\alpha_{\ell,\text{exact}}$ that is not captured by the asymptotic form.

\subsubsection{Behavior of \texorpdfstring{$\alpha_\ell$}{αℓ} for \texorpdfstring{$\mu \rightarrow 0$}{μ→0}}
\label{sub:mu_zero}

To understand why the large $\mu$ limit of $\alpha_\ell$ fails as $\mu \rightarrow 0$,
let us consider the exact behavior of $\alpha_\ell(\mu)$ for small $\mu$.
Expanding $w(r,\mu)$, Eq.~\eqref{eq:w} for $\mu \rightarrow 0$, we get
\andreas{
\begin{equation}
    \label{eq:w-small}
    w(r,\mu) = \frac{2}{\sqrt{\pi}} \mu + \mathcal{O}(\mu^{3} r^2)
\end{equation}
With errors to order $\mu^{3}$} (when the expectation value of $r^2$ does not diverge), the expectation value of $\bar{\op{W}}$ and the energy are shifted for all $\ell$,
\begin{align}
    \expect{\bar{\op{W}}(\mu)}{\Psi(\mu)}  & \rightarrow  \expect{\bar{\op{W}}(0)}{\Psi(0)}  - \frac{N(N-1)}{\sqrt{\pi}} \mu + \dots \\
    E(\mu) & \rightarrow E(0) + \frac{N(N-1)}{\sqrt{\pi}} \mu + \dots
\end{align}
Thus, using Eq.~\eqref{eq:alpha-exact},
\begin{align}
    \alpha_\ell(\mu \rightarrow 0) & = \frac{E-E(\mu)}{\expect{\bar{\op{W}}(\mu)}{\Psi(\mu)}} \\
    \label{eq:alpha-small}
 &\rightarrow  \frac{E-E(0)}{\expect{\bar{\op{W}}(0)}{\Psi(0)}} + \frac{N(N-1)}{\sqrt{\pi}} \frac{E-E(0)-\expect{\bar{\op{W}}(0)}{\Psi(0)}}{\expect{\bar{\op{W}}(0)}{\Psi(0)}^2} \, \mu + \dots
\end{align}
As $E \le E_\op{H}$, the slope of $\alpha_\ell(\mu \rightarrow 0)$ decreases with $\mu$, while for $\mu \rightarrow \infty$, $\alpha_\ell$ increases with $\mu$.

\subsubsection{Errors produced by the cutoff at \texorpdfstring{$L$}{L}}

\begin{figure}
    \centering
    \includegraphics[width=0.8\columnwidth]{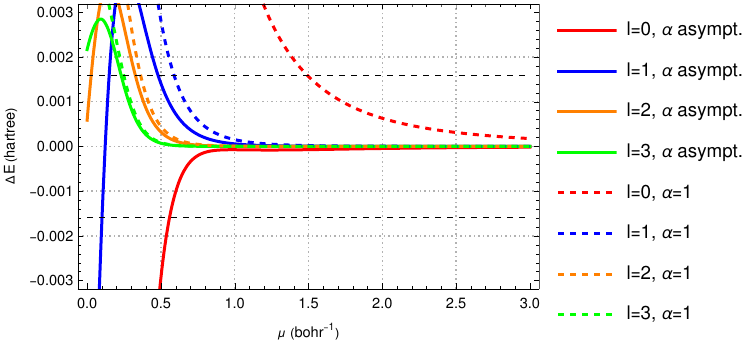}
    \includegraphics[width=0.8\columnwidth]{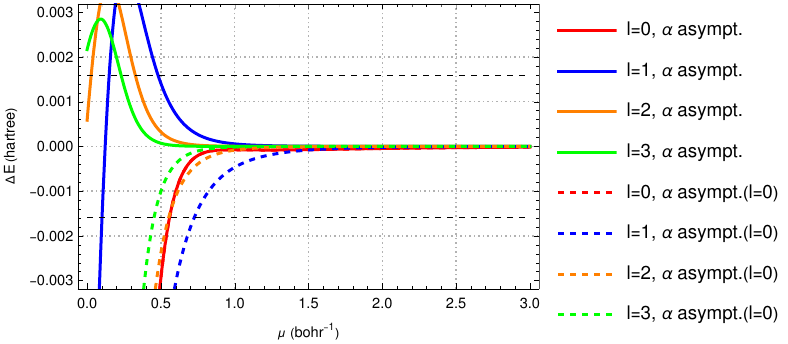}
    \includegraphics[width=0.8\columnwidth]{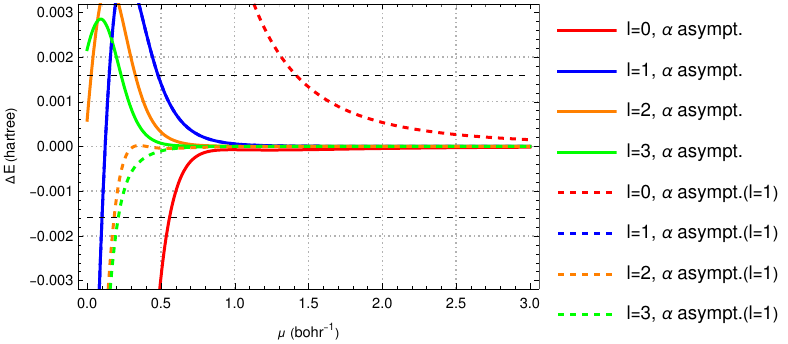}
    \caption{Energy errors as function of the model parameter, $\mu$ for the lowest energy state for $\ell=0$ (red curves), $\ell=1$ (blue curves), $\ell=2$ (orange curves), and $\ell=3$ (green curves). The full curves show the results using the correct $\alpha_\ell$. In the top panel, the dashed curves show the errors produced using $\alpha=1$, {\em i.e.}, while the middle panel shows those produce by using $\alpha_0(\mu)$, and the bottom panel those produced by using $\alpha_1(\mu)$. The horizontal dashed lines indicate the range of chemical accuracy.}
    \label{fig:dewith}
\end{figure}
In order to study the importance of using a cutoff in $L$, Eqs.~\eqref{eq:ac-0} and ~\eqref{eq:ac-1}, let us again consider harmonium for the states of lowest energy with given $\ell=0,1,2,3$.
We compare in fig.~\ref{fig:dewith} the errors of
$E(\mu) + \alpha_\ell(\mu) \expect{\bar{\op{W}}(\mu)}{\Psi(\mu)} $
with those produced when we replace the large $\mu$ form of $\alpha_\ell$, Eq.~\eqref{eq:alpha} by 1 (that is, use $E\approx E_\op{H}(\mu)$), or that corresponding to a different $\ell$.
We notice that using the incorrect $\alpha_\ell$ brings larger errors for $\ell=0$ or $\ell=1$, but not for the higher values of $\ell$.
The errors are showing up at small $\mu$, because of using the form of $\alpha$ valid only for large $\ell$.
In the top panel we compare the results using $\alpha_\ell$, Eq.~\eqref{eq:alpha} with those obtained using a $\mu$-independent value, $\alpha=1$, that is, using the expectation value of the Hamiltonian.
There is a significant effect for $\ell=0$, but the effect is reduced when increasing $\ell$.
This is consistent with the result shown in fig.~\ref{fig:alphas} where $\alpha_\ell$ approaches 1 as $\ell$ increases.
The middle panel shows the same curves with $\alpha_\ell(\mu)$ but compares them with the results obtained by using $\alpha_0(\mu)$ for all states.
Of course, there is no change for the $\ell=0$ state, and we see some worsening for $\ell > 0$, especially for $\ell=1$.
The bottom panel makes now the comparison with the results obtained by using $\alpha_1(\mu)$.
Of course, there is no change for the $\ell=1$ state, but there is some significant worsening for $\ell = 0$.
Strangely, there is even some improvement for $\ell=2$ in a region around $\mu=0.3$.
The results for harmonium thus tend to support the limitation of the sum over $\ell$ to $L=1$, as proposed in Eqs.~\eqref{eq:wbars} and \eqref{eq:wbart}.

\subsection{Total energy errors}
We analyze in this section the errors $E_{L=0}(\mu)-E(\mu=\infty)$ and $E_{L=1}(\mu)-E(\mu=\infty)$ produced within the large basis set for $2 \le N \le 6$.

\begin{table}[]
    \centering
    \begin{tabular}{cc S[table-format=7] S[table-format=2.5] S[table-format=2.5] S[table-format=2.5]}
 \hline
 $N$ & State     & \mc{Number of determinants} &  \mc{$E_{\text{exFCI}}$} &  \mc{$E_{\text{CI}}$} & \mc{Reference}  \\
 \hline
  2  & $^1S$   &         4364             &     2.00018    &     2.00018   & 2.00000\cite{Kin-TCA-96}   \\
     & $^3P$   &         3912             &     2.35967    &     2.35967   & 2.35966\cite{Kin-TCA-96}   \\
     & $^1P$   &         4482             &     2.50028    &     2.50028   & 2.50000\cite{Kin-TCA-96}      \\
 \hline
  3  & $^2P$     &      52833             &     4.01355    &     4.01355   & 4.01322\cite{CioStrMat-JCP-12} \\
     & $^4P$     &      22048             &     4.31073    &     4.31073   & 4.31069\cite{CioStrMat-JCP-12} \\
     & $^2D$     &      30025             &     4.36665    &     4.36665   & 4.36639\cite{VarNavUsuSuz-PRB-01}   \\
 \hline
  4  & $^3P$     &      82929             &     6.34939    &     6.34939   & 6.34883\cite{CioStrMat-JCP-14} \\
     & $^1D$     &     165444             &     6.38642    &     6.38643   & 6.38554\cite{CioStrMat-JCP-14} \\
     & $^1S$     &     189780             &     6.44643    &     6.44644   & 6.44532\cite{CioStr-JCP-17}  \\
     & $^5S$     &      75842             &     6.58730    &     6.58730   & 6.58719\cite{CioStrMat-JCP-14} \\
  \hline
  5  & $^4S$     &    1797043             &     8.99575    &     8.99577   & 8.99484\cite{CioStr-JCP-18} \\
     & $^2D$     &     568582             &     9.05031    &     9.05034   & 9.04883\cite{CioStr-JCP-18} \\
     & $^2P$     &    1036673             &     9.08937    &     9.08938   & 9.08775\cite{CioStr-JCP-18} \\
  \hline
  6  & $^3P$     &    5148761             &    12.03326    &    12.03328   &  12.03068\cite{Str-JCP-16} \\
     & $^1D$     &    7123185             &    12.06863    &    12.06865   &  12.06540\cite{Str-JCP-16} \\
 \hline
    \end{tabular}
    \caption{Extrapolated FCI energy ($E_{\text{exFCI}}$), CI energy ($E_{\text{CI}}$) and number of Slater determinants at $\omega=1/2$ and $\mu=\infty$. Values from the literature are used as a reference.}
    \label{tab:energies}
\end{table}

In Table~\ref{tab:energies}, we present the estimated FCI energy obtained by extrapolating to $E_{\text{PT2}}\to 0$ from the CIPSI calculation ($E_{\text{exFCI}}$), and the variational energy associated with $\Psi(\mu=\infty)$, denoted as $E_{\text{CI}}$, in comparison with the most accurate values we found in the literature.\cite{Kin-TCA-96,CioStrMat-JCP-12,VarNavUsuSuz-PRB-01,CioStrMat-JCP-14,CioStr-JCP-17,CioStr-JCP-18,Str-JCP-16}.


We plot the errors $E_{L=0}(\mu)-E(\mu=\infty)$ and $E_{L=1}(\mu)-E(\mu=\infty)$ as functions of $\mu$, usually in the range around chemical accuracy.
All generated plots share a notable characteristic: the accuracy of the approximations significantly deteriorates when the parameter $\mu$ is small.
This outcome aligns with expectations, as the approximations are designed to become exact in the limit of large $\mu$, as discussed in subsection~\ref{sub:mu_zero}.
Another recurrent observation is that the model energies, $E(\mu)$, are poor estimates of the exact energy $E$, particularly in scenarios involving paired electrons.
In certain cases, the magnitude of these errors is so large that $E(\mu)$ doesn't appear in the plots.
In contrast, $E_\op{H}$, presents a marked improvement over $E(\mu)$ by incorporating a first-order correction to $E(\mu)$.
Moreover, it is consistently observed that the difference $E_\op{H}(\mu) - E$ is always non-negative.

\subsubsection{Two electrons}

\begin{figure}
    \centering
    \includegraphics[width=0.8\columnwidth]{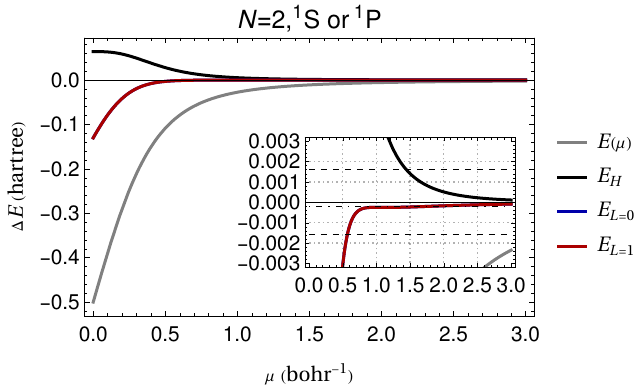}
    \includegraphics[width=0.8\columnwidth]{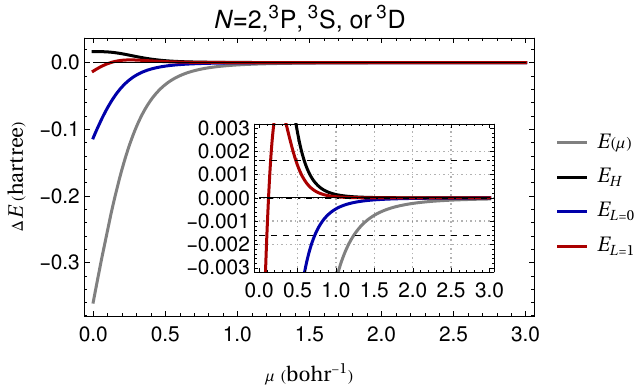}
    \caption{Energy errors (in \SI{}{\milli\hartree}) for Harmonium, $\omega=1/2$, $N=2$, ground state singlet ($^1S$, top panel), and first excited triplet state ($^3P$, bottom panel). Error of the model energy, $E(\mu)$, gray curve; $E_\op{H}(\mu)$, Eq.~\eqref{eq:var}, black curve;  of $E_{L=0}(\mu)$, Eq.~\eqref{eq:ac-0}, blue curve, covered by that of $E_{L=1}(\mu)$, Eq.~\eqref{eq:ac-1}, red curve. The inset zooms into the region of chemical accuracy, marked by horizontal dashed lines.  The horizontal dot-dashed line shows the difference between the most accurate value in the literature energy and the FCI estimate of $E$ in the basis set we use.}
    \label{fig:de-harm-2}
\end{figure}

Due to the separability in this particular case, other states with the same $\ell$ present the same errors, {\em e.g.}, the first excited $^1P_u$, the third $^1S_g$ state presents the same error as the $^1S_g$ ground state, or the first excited $^3P_u,^3S_g, ^3P_g, ^3D_g$ have all the same error.
(The first excited $^1D_g$, not shown here, is a non-natural singlet, as it has $\ell=2$.)
We see in fig.~\ref{fig:de-harm-2} that the errors due to the basis set are very small.
(They are very close to values produced on a grid, see ref.~\citenum{SavKar-JCP-23}.)
For the singlet states, $E_\op{H}(\mu)$ is within chemical accuracy only for $\mu > 1.5$~bohr$^{-1}$.
Using $E_{L=0}(\mu)$ significantly improves the range of accuracy in the singlet case.
Using $E_{L=1}(\mu)$ has no effect on the singlet state, as there is only one singlet electron pair: the $E_{L=0}(\mu)$ and $E_{L=1}(\mu)$ curves are superimposed.
For the triplet state,  fig.~\ref{fig:de-harm-2} (bottom), there is a difference between $E_{L=0}(\mu)$ and $E_{L=1}(\mu)$.
In this case, $\langle \bar{\op{W}} \rangle_s =0$,  and $\langle \bar{\op{W}} \rangle_t =\langle \bar{\op{W}} \rangle$.
However, we note that the error made by $E_{L=0}(\mu)$ is not very large.
Furthermore, for the triplet state, the improvement over $E_\op{H}$ is minimal, as electrons are already kept apart, as discussed above.
We do not show here results for non-natural states ($\ell>1$), for example the second $^1S$ state, but they can be found in ref.~\citenum{SavKar-JCP-23}.
One can see there that the trend seen for triplets is further enhanced for $\ell > 1$, singlets, or triplets.

\subsubsection{Three electrons}

\begin{figure}
    \centering
    \includegraphics[width=0.75\columnwidth]{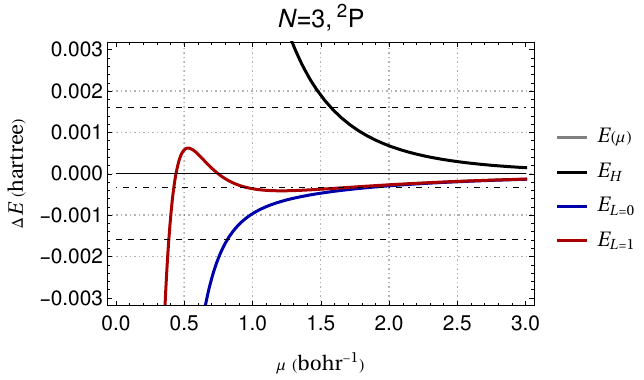}
    \includegraphics[width=0.75\columnwidth]{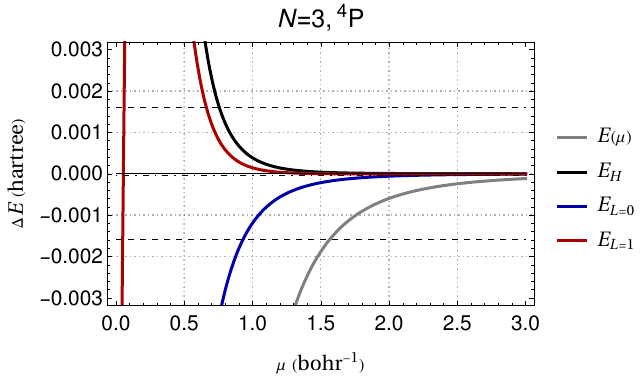}
    \includegraphics[width=0.75\columnwidth]{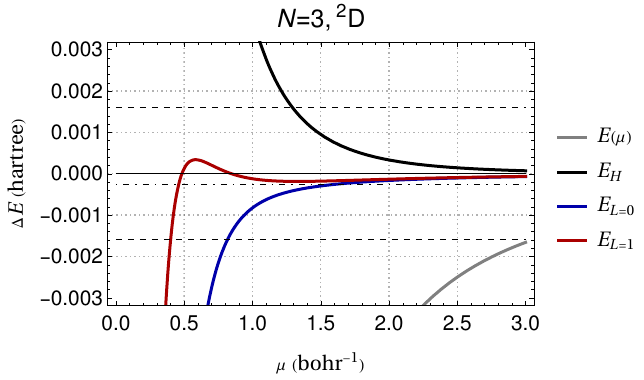}
    \caption{Energy errors (in \SI{}{\milli\hartree}) for Harmonium, $\omega=1/2$, $N=3$. Top panel: $^2P$, central panel: $^4P$, bottom panel: $^2D$. Color codes as in fig.~\ref{fig:de-harm-2}. \andreas{The curve of $E(\mu)$ for the $^2P$ state is outside of the plotting window.}}
    \label{fig:de-harm-3}
\end{figure}

For $N=3$, we consider the lowest three states.
The ground state $^2P$ has a dominant s$^2$p configuration, while the two excited states ($^4P, ^2D)$ correspond to a high- and low-spin configuration sp$^2$.
The $^4P$ presents curves with features similar to those for the $N=2,~^3P$.
The advantage of of using $E_{L=1}$ becomes clear for the doublet states, where both singlet and triplet pairs are present: it lowers the value of $\mu$ for which chemical accuracy is reached.

\subsubsection{Four electrons}
\begin{figure}
    \centering
    \includegraphics[height=0.20\textheight]{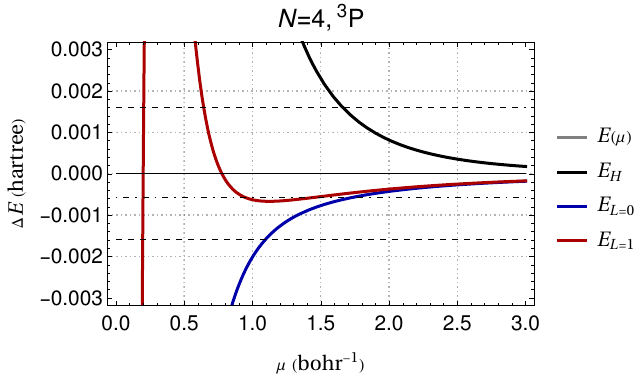} \\
    \includegraphics[height=0.20\textheight]{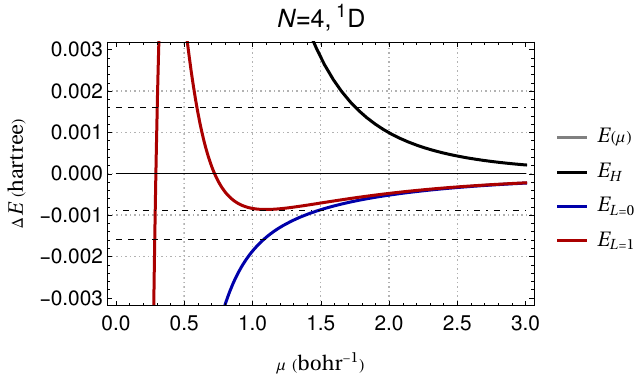} \\
    \includegraphics[height=0.20\textheight]{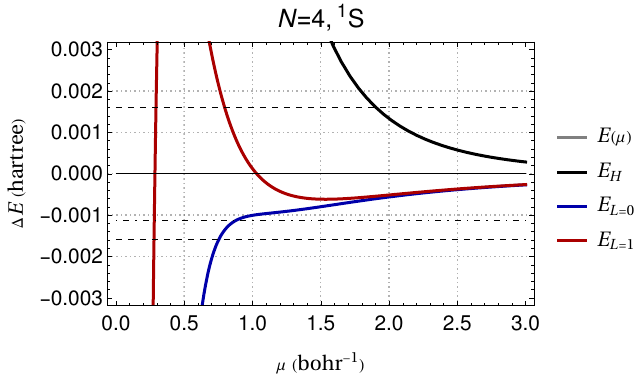}\\
    \includegraphics[height=0.20\textheight]{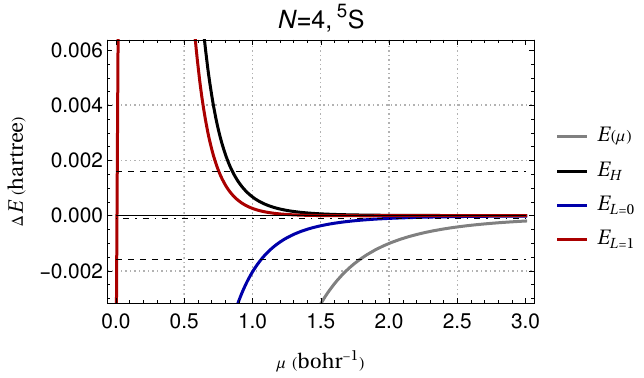}
    \caption{Energy errors (in \SI{}{\milli\hartree}) for Harmonium, $\omega=1/2$, $N=4$. From top to bottom: $^3P, ^1D, ^1S, ^5S$. Color codes as in fig.~\ref{fig:de-harm-2}. \andreas{Except for the $^5S$ state, the curves of $E(\mu)$ is outside of the plotting window.}}
    \label{fig:de-harm-4}
\end{figure}

For $N=4$ we consider the states dominated by an s$^2$p$^2$ configuration (the ground state, $^3P$, and the excited $^1D$ and $^1S$ states), and the $^5S$ state, dominated by the sp$^3$ configuration (fig.~\ref{fig:de-harm-4}).
Again, there is no surprising effect for the state where all spins are parallel ($^5S$): the curves are similar to those seen for $N=2$ or $N=3$.
However, we start seeing clearly a new effect for the other states.

For systems with $N=2$ or $N=3$ electrons, the energies $E_{L=1}(\mu)$ exhibit poor accuracy at low values of $\mu$.
Nevertheless, as $\mu$ increases, a convergence point is reached where the curves approximate the exact energy, obtained from a complete basis set, and then remain flat beyond this point.
This pattern of behavior is also observed for $N=4$ in the $^5S$ state.
However, for other states, the plateau effect is absent.
Instead, there exists a specific $\mu$ value at which $E_{L=1}(\mu)$ approaches the complete basis set estimate.
Beyond this value, a gradual, albeit slow, increase in error occurs as $\mu$ continues to increase, attributable to the basis set's inability to fully capture the short-range correlation effects.

For states with higher spin, the basis set proves sufficiently accurate, yielding minimal error when $\mu$ approaches infinity.
Conversely, for states with lower spin, the basis set incurs an approximate error of $1/2$~kcal/mol at $\mu=\infty$.
As the value of $\mu$ is reduced, the results appear to align more closely with the exact results, effectively mitigating the basis set error.
This improvement, however, begins to compete with the limitations inherent to approximations that are primarily valid in the regime of large $\mu$.


\subsubsection{Five and six electrons}

\begin{figure}
    \centering
    \includegraphics[width=0.75\columnwidth]{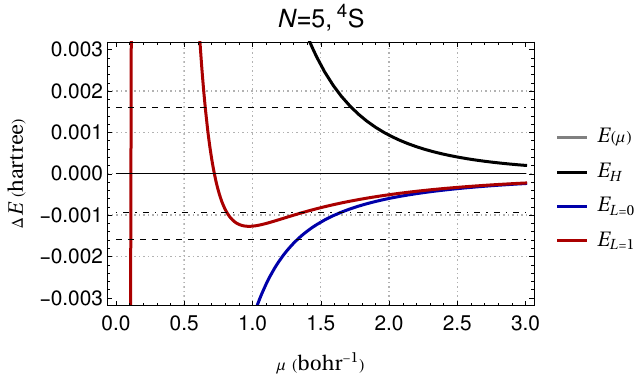}
    \includegraphics[width=0.75\columnwidth]{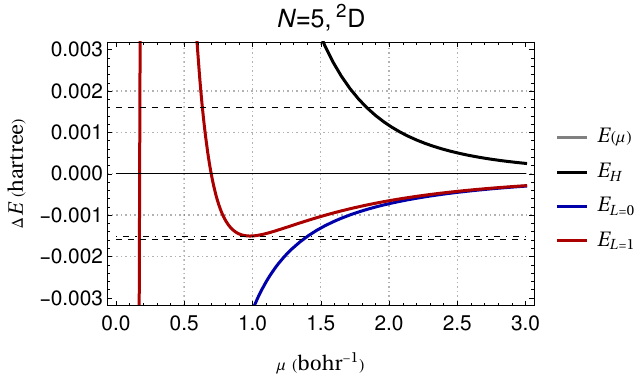}
    \includegraphics[width=0.75\columnwidth]{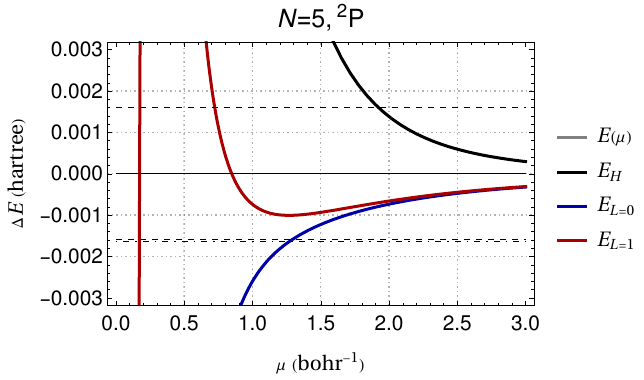}
    \caption{Energy errors (in \SI{}{\milli\hartree}) for Harmonium, $\omega=1/2$, $N=5$. From top to bottom: $^4S, ^2D, ^2S$. Color codes as in fig.~\ref{fig:de-harm-2}. \andreas{The curves of $E(\mu)$ are outside of the plotting window.}}
    \label{fig:de-harm-5}
\end{figure}

\begin{figure}
    \centering
    \includegraphics[width=0.8\columnwidth]{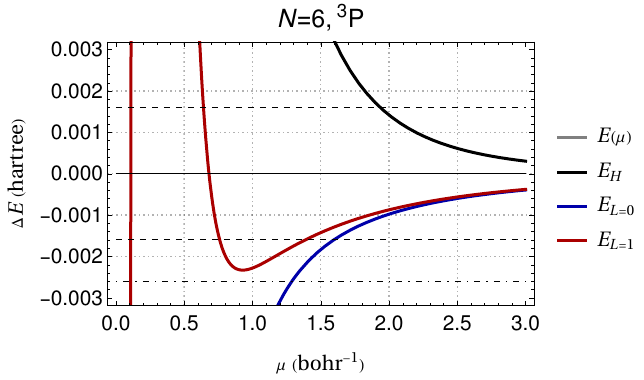}
    \includegraphics[width=0.8\columnwidth]{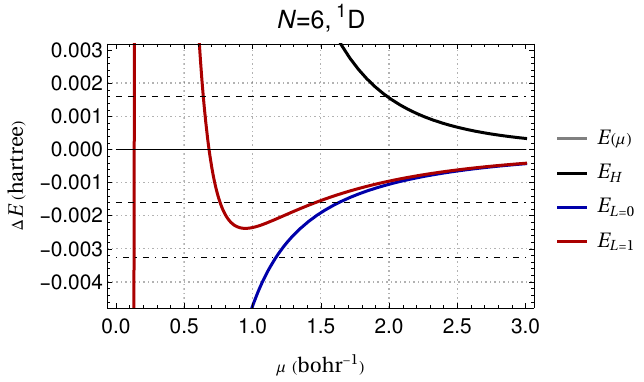}
    \caption{Energy errors (in \SI{}{\milli\hartree}) for Harmonium, $\omega=1/2$, $N=6$. From top to bottom: $^3P, ^1D$,
    Color codes as in fig.~\ref{fig:de-harm-2}. Note the different scale for the $^1D$ system. \andreas{The curves of $E(\mu)$ are outside of the plotting window.}}
    \label{fig:de-harm-6}
\end{figure}


For $N=5$, as shown in Fig.~\ref{fig:de-harm-5}, and $N=6$, depicted in Fig.\ref{fig:de-harm-6}, an improvement is observed across all states by employing $E_{L=1}(\mu)$ over $E_{L=0}(\mu)$.
This improvement follows patterns similar to those previously identified.
It is notable that the basis set errors in the FCI calculations are larger than for $N=4$, reaching 1-2 kcal/mol.
Consequently, the benefits of selecting a lower $\mu$ value become more pronounced, as illustrated in Fig.~\ref{fig:de-harm-6}.
At a $\mu$ value of approximately 0.9~bohr$^{-1}$, the difference to the FCI energy in the chosen basis set surpasses the threshold of chemical accuracy.
Nonetheless, when compared to the most accurate calculations available in the literature\cite{Str-JCP-16}, these errors remain within the bounds of chemical accuracy.

In all examples shown, the errors increase rapidly as $\mu$ drops below a value lying between 0.5 and 1.0~bohr$^{-1}$.
This value depends on the system and the state.
To be safe, one would prefer to choose a larger value for $\mu$.
However, this does not only make the calculation more costly, but, in order to have a small basis set error, it is preferable to choose a small $\mu$.

Not treated in this paper are more compact, or more diffuse systems.
This can be analyzed in a similar way, by changing $\omega$.~\cite{KarSav-PTCP-23}
Fortunately, one is often interested in properties describing changes in the outer valence shell, and in such cases, smaller values of $\mu$ are sufficient.

\subsection{Errors in energy differences}
\begin{figure}
    \centering
    \includegraphics{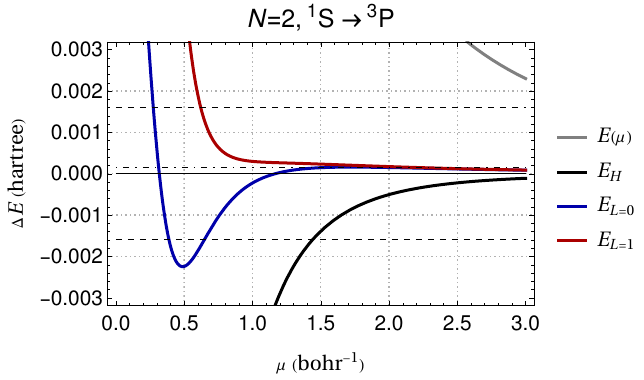}
    \includegraphics{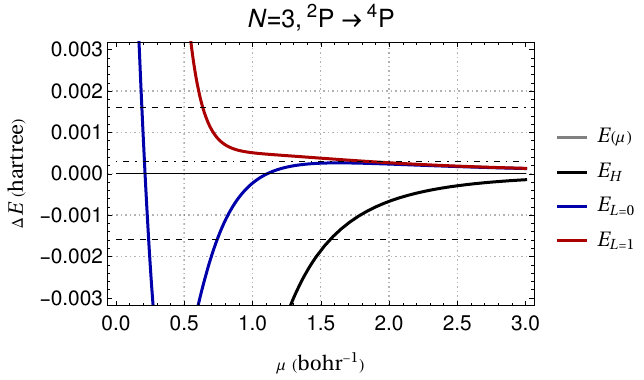}
    \includegraphics{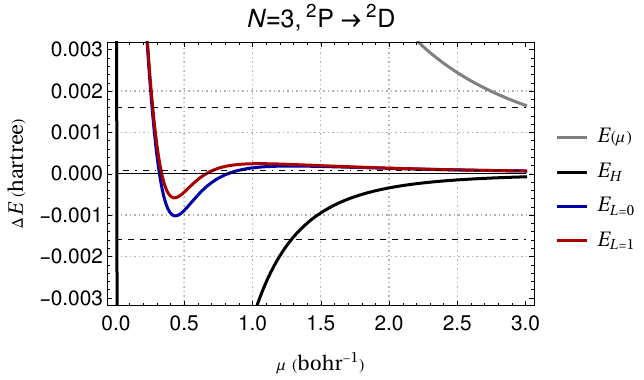}
    \caption{Errors in excitation energies, $N=2$, and $N=3$.
    The transitions are indicated above the plots. The color code and the horizontal lines have the same meaning as in fig.~\ref{fig:de-harm-2}. \andreas{For the ${^2P} \rightarrow {^4P}$ transition, the curve of $E(\mu)$ is outside of the plotting window.}}
    \label{fig:excit-2-3}
\end{figure}

\begin{figure}
    \centering
    \includegraphics{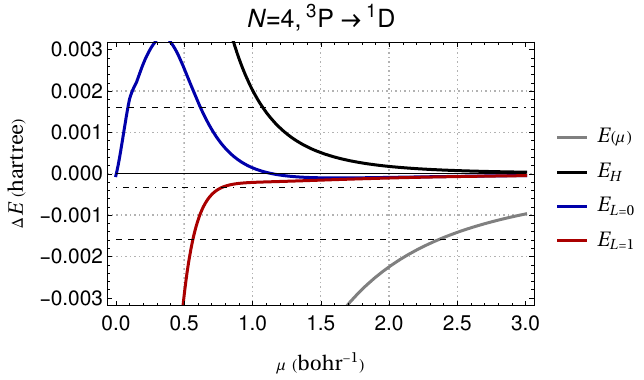}
    \includegraphics{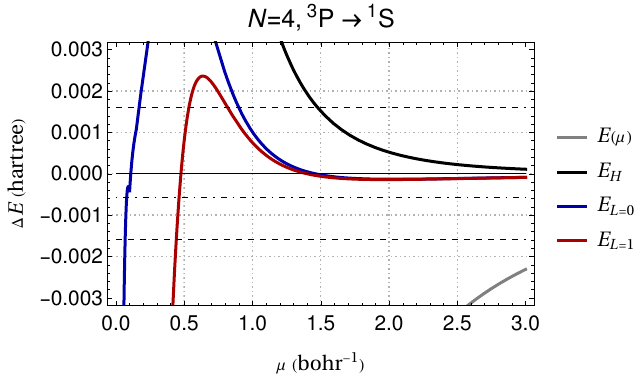}
    \includegraphics{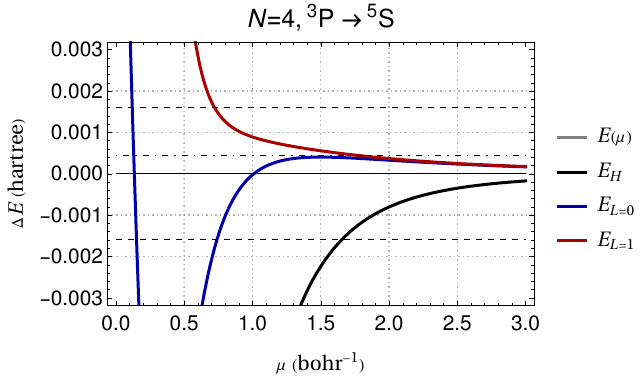}
    \caption{Errors in excitation energies, $N=4$. The transitions are indicated above the plots. The color code and the horizontal lines have the same meaning  as in fig.~\ref{fig:de-harm-2}. \andreas{For the ${^3P} \rightarrow {^5S}$ transition, the curve of $E(\mu)$ is outside of the plotting window.}}
    \label{fig:excit-4}
\end{figure}

\begin{figure}
    \centering
    \includegraphics{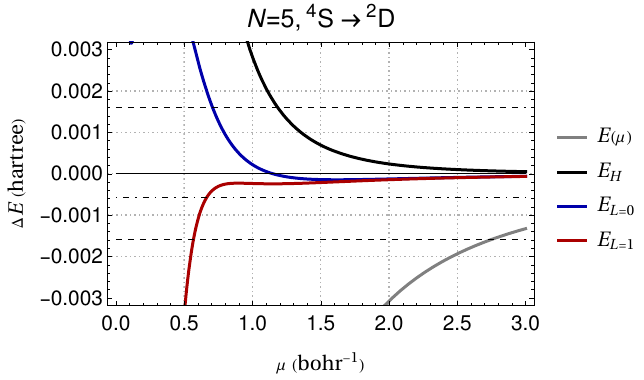}
    \includegraphics{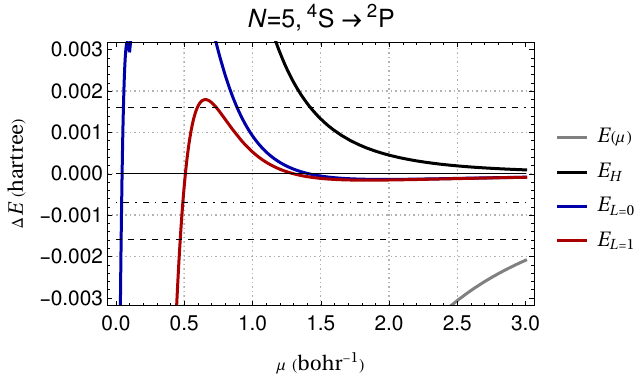}
    \includegraphics{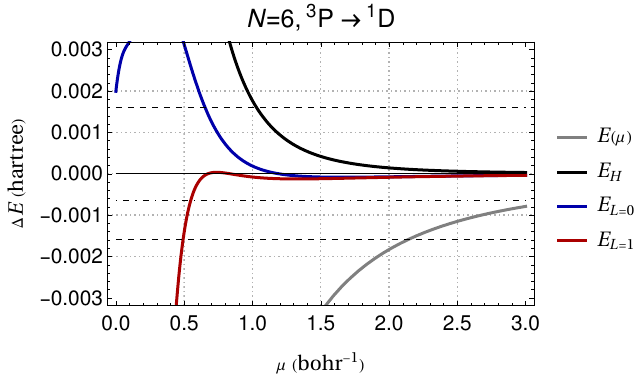}
    \caption{Errors in excitation energies, $N=5$ and $N=6$. The transitions are indicated above the plots. The color code and the horizontal lines have the same meaning as in fig.~\ref{fig:de-harm-2}.}
    \label{fig:excit-5-6}
\end{figure}


Typically, there is a greater interest in energy differences than in absolute total energies.
It is often observed that due to error compensation, results for energy differences are more accurate than those for total energies.
For instance, in the case of Harmonium with two electrons, the errors for systems possessing the same quantum number $\ell$ are identical, leading to exact error cancellation.
However, this exact compensation does not hold for excitation energies between states with different $\ell$, such as the transition from $^1S$ to $^3P$ for $N=2$, as illustrated in Fig.~\ref{fig:excit-2-3}.
Although exact error cancellation may not occur, it is possible that the range of $\mu$ values within which errors remain minimal could be broader, for example, in the $^2P \rightarrow ^2D$ excitation for $N=3$, shown in Fig.~\ref{fig:excit-2-3}.

\begin{figure}
    \centering
    \includegraphics[width=0.45\textwidth]{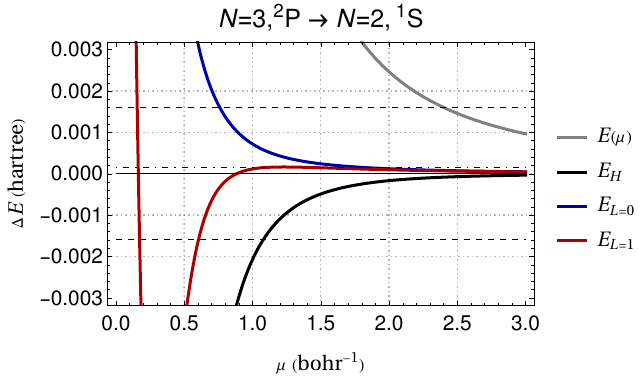}
    \includegraphics[width=0.45\textwidth]{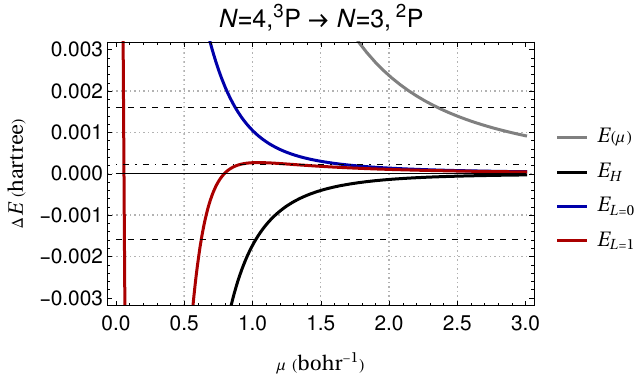}
    \\ \vspace{3mm}
    \includegraphics[width=0.45\textwidth]{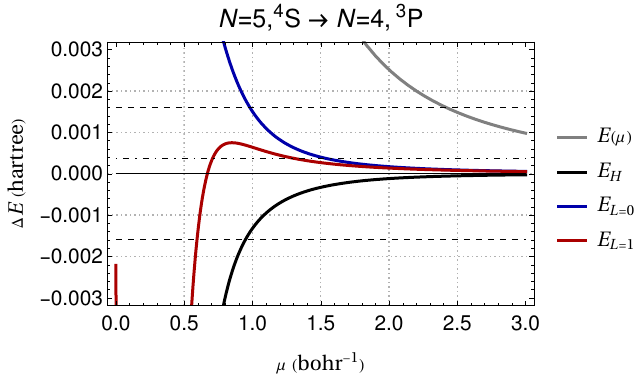}
    \includegraphics[width=0.45\textwidth]{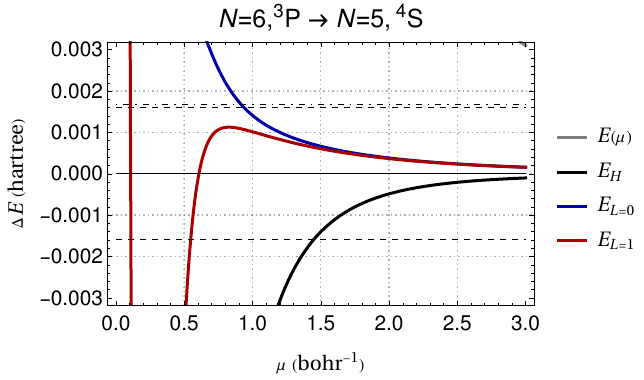}
    \caption{Errors of "ionization energies". The numbers of electrons and states are given above the plots. The color code and the horizontal lines have the same meaning as in fig.~\ref{fig:de-harm-2}. \andreas{For the ionization of the $N=6$ ground state, the curve of $E(\mu)$ is outside of the plotting window.}}
    \label{fig:ip}
\end{figure}

Overall, as evidenced in Figs.~\ref{fig:excit-2-3}-\ref{fig:ip}, energy differences stay reliable down to approximately $\mu = 0.5$ bohr$^{-1}$. Beyond this, particularly at lower $\mu$ values, the substantial deterioration in accuracy renders the excitation energies unreliable.

%
%

The results for the total energy of $N=2$ using $E_{L=1}$ proved to be superior to those obtained with $E_{L=0}$ for the $^3P$ state, whereas the choice between $L=0$ and $L=1$ did not affect the results for the $^1S$ state.
Interestingly, around $\mu \approx 0.5$ bohr$^{-1}$, an error compensation phenomenon is observed, making $E_{L=0}$ marginally better.

It is evident across all scenarios that both $E_{L=0}$ and $E_{L=1}$ not only surpass the model energies but also offer more accurate excitation energies than those derived from $E_\op{H}$.
It's important to note that $E_\op{H}$ does not provide a bound for energy differences.
As illustrated in Fig.~\ref{fig:excit-4}, the excitation energy calculated using $E_\op{H}$ can vary, being either higher or lower than the precise value depending on the specific excitation being examined.
This variability is also seen in the FCI calculations.
Nonetheless, a certain level of error compensation occurs due to the basis set incompleteness, which tends to shift both states to higher energies.

This pattern of observations regarding excitation energies and their relative accuracy extends to systems with $N=4$, $N=5$, or $N=6$ electrons, as demonstrated in Figs.~\ref{fig:excit-4} and \ref{fig:excit-5-6}.

We now consider energy differences between states with different number of electrons.
We call the difference between the energy of the ground state with $N-1$ electrons and that with $N$ electrons "ionization potential", although electrons cannot escape the potential given in Eq.~\eqref{eq:V}.
We see in fig.~\ref{fig:ip} that the errors of the "ionization potentials" follow the trends mentioned for the excitation energies, and do not need a further discussion.

\begin{figure}
    \centering
    \includegraphics[width=0.40\textwidth]{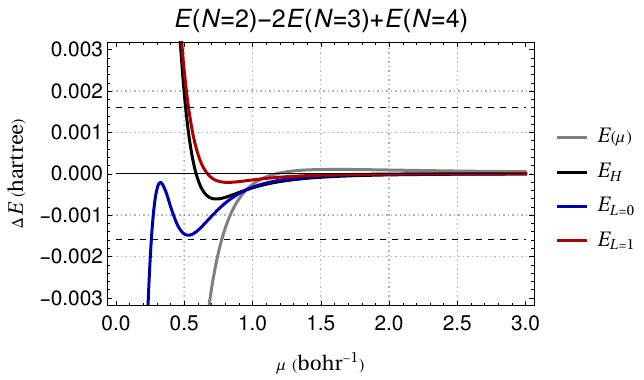}
    \includegraphics[width=0.40\textwidth]{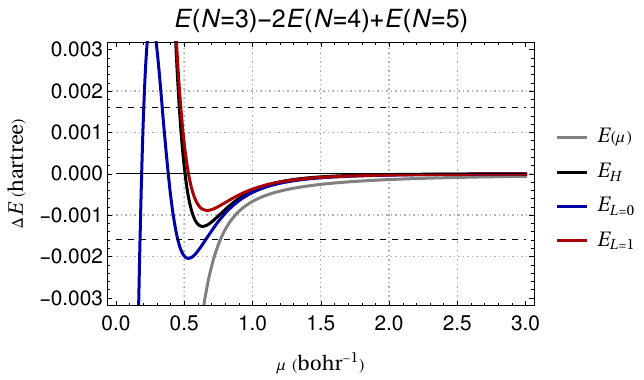}
    \includegraphics[width=0.40\textwidth]{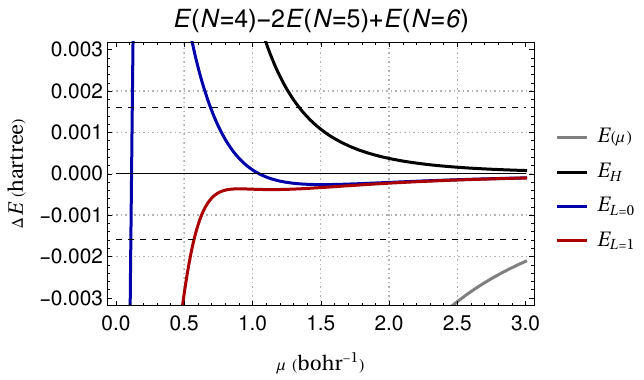}
    \caption{Errors of "band gaps". The numbers of electrons and states are given above the plots. The color code and the horizontal lines have the same meaning as in fig.~\ref{fig:de-harm-2}.}
    \label{fig:gap}
\end{figure}
We can also look at "band gaps", $E(N-1)-2E(N)+E(N+1)$, from the ground state energies for systems with different $N$, fig.~\ref{fig:gap}.
We see that in two of the three cases considered, $N=3$, and $N=4$, $E_\op{H}$ is much better than we observed for excitation energies.
However, this observation cannot be made for $N=5$.
It is much safer to compute with the same computational effort $E_{L=0}$.

\section{Summary and perspectives}
\label{summary}

This papers proposes $E_{L=1}$, eq.~\eqref{eq:ac-1}, to approximate ground state energies.
The computational effort is not much larger than that needed to compute expectation values of the Hamiltonian, $E_\op{H}$, eq.~\eqref{eq:var} while we obtain a significant improvement.
The method requires first solving accurately the Schr\"odinger equation with a model, long-range operator for the interaction between electrons.
A system-independent correction is applied.
It does not use any empirical parameter, and becomes exact as the physical (Coulomb) interaction is approached (is valid when the parameter $\mu$ characterizing the model is large).
It uses system-independent prefactors $\alpha_0$ and $\alpha_1$ to the singlet and triplet components of the repulsion between the electrons.
They depend on the range-separation parameter, $\mu$, but not on the system or state under consideration; they are universal (in the language of density functional theory).

\begin{figure}
    \centering
    \includegraphics[width=0.75\textwidth]{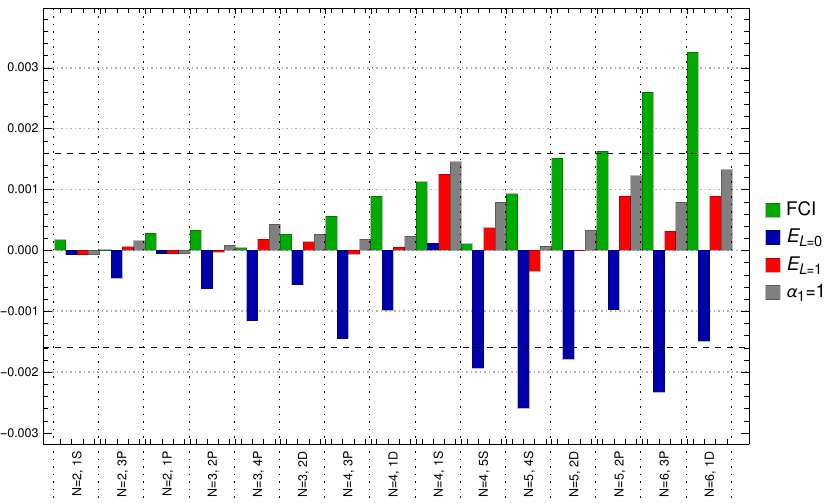}
    \caption{Total energy errors in $E_{\text{h}}$ (w.r.t. the best estimate from literature) for $\mu=1$ with different methods (FCI, green bars;  $E_{L=0}$, blue bars; $E_{L=1}$, red bars; with $\alpha_1=1$, gray bars; the horizontal dashed lines indicate the domain of chemical accuracy ($\pm 1$~kcal/mol).}
    \label{fig:de-at-mu-1}
\end{figure}
Numerical results are obtained for $N=2$ to $6$  electrons in a quadratic confinement.
\andreas{Although we have treated in this paper only systems with spherical symmetry, we would like to stress that the method presented is generally applicable.
The results }
are summarized in fig.~\ref{fig:de-at-mu-1} showing the errors in the total energies.
As the method is approaching the FCI result as $
\mu \rightarrow \infty$, but the basis set effects diminish with $\mu$, we present in this figure the results for a compromise value, $\mu=1$.
The errors of the model energies, $E(\mu)$ and $E_\op{H}(\mu)$ are not shown, being much larger in all cases.
We see that the errors are, for all systems studied, within chemical accuracy, and very often smaller that the FCI calculation used to generate them.

The errors in energy differences sometimes are often better than those of the total energies through compensation.
Although there is no guarantee for improvement, we find in general some improvement.
For example, the  $^4S \rightarrow ^2D$ excitation energy for five electrons, fig.~(\ref{fig:excit-2-3}) is within chemical accuracy down to $\mu \approx 0.4$~bohr$^{-1}$ while the the total energies are within chemical accuracy only down to $\mu \approx 0.7$~bohr$^{-1}$.
However, one can consider the improvement for energy differences modest.
The explanation is that large $\mu$ dependence of the prefactors $\alpha_\ell$  is different at small $\mu$ from that at large $\mu$, fig.~\ref{fig:alphas} and eq.~\eqref{eq:alpha-small}.
This brings a rapid worsening of the results as $\mu$ diminishes, and this worsening is reflected also in energy differences: there is a change of regime in $\alpha_\ell$.
Furthermore, we would like to point out that the FCI property of providing  exact upper bounds to the energies is lost with $E_{L=1}$.
However, even for FCI, the bounding property is lost for energy differences.

Fig.~\ref{fig:de-at-mu-1} shows also the effect of modifying the corrective prefactor $\alpha_1$, eq.~\eqref{eq:alpha-1-mu}.
It lies between $\alpha_0$, eq.~\eqref{eq:alpha-0-mu} and 1, eq.~\eqref{eq:alpha-ineq}.
Choosing the lower bound corresponds to treat triplet electron pairs (often called parallel electron pairs) the same way as singlet electron pairs (often called anti-parallel electron pairs).
In choosing the upper limit (\andreas{$\alpha_1 = 1$}) one ignores corrections to $E_\op{H}$ for triplet electron pairs.
The bounds to $\alpha_1$ also bounds $E_{L=1}$, see eq.~\eqref{eq:e-ineq}.
We see in fig.~\ref{fig:de-at-mu-1} that the difference between them can give an idea about the error of $E_{L=1}$ without any additional computational effort.
This estimate can be understood seen as a detector for the change of regime.
In the same way, one can also consider $E_{L=1}-E_{L=0}$ as a measure of accuracy.
However, these estimators of the accuracy are not guaranteed.
In the case when the triplet component of the repulsion between electrons vanishes these estimators are 0.
This can be also the case for energy differences where the triplet components compensate, such as
the $^2P \rightarrow ^2D$ excitation energy, for $N=3$, fig.~\ref{fig:excit-2-3}.

A few important problems remain to be studied, but this will become the object of a different paper.
Here we only mention them.
The most important is the choice of the model system, that is, of the parameter $\mu$.
It has the dimension of an inverse distance.
In a compact system, the change of regime occurs at larger values of $\mu$ than for diffuse systems.
This impacts the correct treatment of size-consistency, the core and the valence part of a given system, etc.
A solution to this problem may be to make $\mu$ locally dependent (see, e.g., refs.~\citenum{PolSavLeiSto-JCP-02, KlaBah-JCTC-20}).
Another important effect is to treat the basis set errors (as can be seen in the FCI errors in fig.~\ref{fig:de-at-mu-1}).
In fact, this can be also treated with a local $\mu$~\cite{GinPraFerAssSavTou-JCP-18,LooPraSceTouGin-JPCL-19,GinSceLooTou-JCP-20,TraGinTou-JCP-23}.

\andreas{
Eq.~\eqref{eq:acE} assumes that the derivative of $E(\lambda,\mu)$ with respect to $\lambda$ exists. However, discontinuities may appear in particular cases, although in our numerical examples we have not noticed such a phenomenon.
A possible explanation for not observing such cases is that we modify only the short range behavior of the wave function, which is universal, as long as $\mu$ is large enough.
When $\mu$ becomes small, the approximations we introduce become less sensible, and a drastic change of the wave function between $\lambda=0$ and $\lambda=1$ could have an additional impact on the accuracy of the results.
Assume now that two eigenvalues $E_i(\lambda,\mu)$ corresponding to eigenfunctions $\Psi_i(\lambda,\mu)$ cross at some value of $\lambda$.
As our method is not limited to the ground state, and only the interaction term is changed along the adiabatic connection, the symmetry of the two states is not changed. Thus, states of different symmetry are allowed to cross.
When an avoided crossing occurs (one is interested in states of the same symmetry), it may be useful to extend the procedure to a matrix formalism, treating both states simultaneously.
}

Another, less important limitation is that through the asymptotic form of the wave function, eq.~\eqref{eq:varphi-asy}, the short-range repulsion is described correctly to orders $\mu^{-2}$ and $\mu^{-3}$ for singlet pairs, and to orders $\mu^{-4}$ and $\mu^{-5}$ for triplet pairs.~\cite{Sav-JCP-20}
Thus, it may be useful to also consider orders $\mu^{-n}, n =3, 4$ for singlet pairs.

\section*{Acknowledgements}
This paper is dedicated to Trygve Helgaker.
We know that he has broad interests, but also likes to penetrate deeply the subjects he touches.
Although it cannot match the clarity of his presentations, we hope nevertheless that our paper might catch his interest, as our viewpoint suggests a different viewpoint on what density functional approximations do, or should do, a subject Trygve was interested in.

This work was realized using HPC resources from CALMIP (Toulouse) under allocation \texttt{p22001}.

\bibliography{biblio}
\end{document}